\documentclass[preprint2]{aastex61}

\newcommand\aastex{AAS\TeX}

\accepted{April 28, 2017}
\shorttitle{\aastex\ OTS44 ALMA}
\shortauthors{Bayo et al.}
\begin{document}

\title{First millimeter detection of the disk around a young, isolated, planetary-mass object}

\correspondingauthor{Amelia Bayo}
\email{ameia.bayo@uv.cl}

\author[0000-0001-7868-7031]{Amelia Bayo}
\affiliation{Inst. F\'isica y Astronom\'ia, Fac. Ciencias, Universidad de Valpara\'iso, Gran Breta\~na 1111, Valpara\'iso, Chile}
%\nocollaboration

%\author{August Muench}
%\affiliation{American Astronomical Society \\
%2000 Florida Ave., NW, Suite 300 \\
%Washington, DC 20009-1231, USA}
%\nocollaboration
%\collaboration{(AAS Journals Data Scientists collaboration)}

\author{Viki Joergens}
\affiliation{Max Planck Institut f\"ur Astronomie, K\"onigstuhl 17, D-69117, Heidelberg, Germany}
%\affiliation{AAS Journals Associate Editor-in-Chief}
%\nocollaboration

\author{Yao Liu}
%\altaffiliation{Creator of AASTeX v6.1}
\affiliation{Max Planck Institut f\"ur Astronomie, K\"onigstuhl 17, D-69117, Heidelberg, Germany}
\affiliation{Purple Mountain Obs., and Key Laboratory for Radio Astronomy,  CAS, 2 West Beijing Road, Nanjing 210008, China}
%\nocollaboration
%\collaboration{(LaTeX collaboration)}

%\author{et al.}
%\affiliation{affiliations}
%%%\affiliation{IOP Publishing, Washington, DC 20005}
%%%\nocollaboration

\author{Robert Brauer}
\affiliation{Institute of Theoretical Physics and Astrophysics, University of Kiel, Leibnizstr. 15, D-24118 Kiel, Germany}

\author{Johan Olofsson}
\affiliation{Inst. F\'isica y Astronom\'ia, Fac. Ciencias, Universidad de Valpara\'iso, Gran Breta\~na 1111, Valpara\'iso, Chile}
\affiliation{Max Planck Institut f\"ur Astronomie, K\"onigstuhl 17, D-69117, Heidelberg, Germany}

\author{Javier Arancibia}
%%\affiliation{AAS Director of Publishing}
\affiliation{Inst. F\'isica y Astronom\'ia, Fac. Ciencias, Universidad de Valpara\'iso, Gran Breta\~na 1111, Valpara\'iso, Chile}
%%\nocollaboration

\author{Paola Pinilla}
\affiliation{Dept. of Astronomy/Steward Obs., University of Arizona, 933 North Cherry Avenue, Tucson, AZ 85721, USA}

\author{Sebastian Wolf}
\affiliation{Institute of Theoretical Physics and Astrophysics, University of Kiel, Leibnizstr. 15, D-24118 Kiel, Germany}

\author{Jan Philipp Ruge}
\affiliation{Institute of Theoretical Physics and Astrophysics, University of Kiel, Leibnizstr. 15, D-24118 Kiel, Germany}

\author{Thomas Henning}
\affiliation{Max Planck Institut f\"ur Astronomie, K\"onigstuhl 17, D-69117, Heidelberg, Germany}

\author{Antonella Natta}
\affiliation{School of Cosmic Physics, Dublin Institute for Advanced Studies, 31 Fitzwilliams Place, Dublin 2, Ireland}
\affiliation{INAF-Osservatorio Astrofisico di Arcetri, L.go E. Fermi 5, I-50125 Firenze, Italy }

\author{Katharine G. Johnston}
\affiliation{School of Physics \& Astronomy, E.C. Stoner Building, The University of Leeds, Leeds LS2 9JT, UK}

\author{Mickael Bonnefoy}
\affiliation{Univ. Grenoble Alpes, CNRS, IPAG, F-38000 Grenoble, France}

\author{Henrik Beuther}
\affiliation{Max Planck Institut f\"ur Astronomie, K\"onigstuhl 17, D-69117, Heidelberg, Germany}

\author{Gael Chauvin}
\affiliation{Unidad Mixta Internacional Franco-Chilena de Astronom\'{i}a, CNRS/INSU UMI 3386 and Departamento de Astronom\'{i}a, Universidad de Chile, Casilla 36-D, Santiago, Chile.}
\affiliation{Univ. Grenoble Alpes, CNRS, IPAG, F-38000 Grenoble, France}



\begin{abstract}
OTS44 is one of only four free-floating planets known to have a disk. We have previously shown that it is the coolest and least massive known free-floating planet ($\sim$12 M$_{\rm Jup}$) with a substantial disk that is actively accreting. We have obtained Band 6 (233 GHz) ALMA continuum data of this very young disk-bearing object. The data shows a clear unresolved detection of the source. We obtained disk-mass estimates via empirical correlations derived for young, higher-mass, central (substellar) objects. The range of values obtained are between 0.07 and 0.63 M$_{\oplus}$ (dust masses). We compare the properties of this unique disk with those recently reported around higher-mass (brown dwarfs) young objects in order to infer constraints on its mechanism of formation. While extreme assumptions on dust temperature yield disk-mass values that could slightly diverge from the general trends found for more massive brown dwarfs, a range of sensible values provide disk masses compatible with a unique scaling relation between $M_{\rm dust}$ and $M_{*}$ through the substellar domain down to planetary masses.
%The first lesson in the tutorial is to remind
%authors that the AAS Journals, the Astrophysical Journal (ApJ), the
%Astrophysical Journal Letters (ApJL), and Astronomical Journal (AJ), all
%have a 250 word limit for the abstract.  If you exceed this length the
%Editorial office will ask you to shorten it.
\end{abstract}



\section{Introduction} \label{sec:intro}

%{\bf Just coming from the proposal slightly edited and updated, but needs to be cut by a lot!!}

Key questions are still unanswered about how low-mass stars come to emerge and gain their masses from their natal molecular clouds, but when we move toward the substellar domain (with masses below the hydrogen-burning mass limit, $\leq$ 0.072$M_{\odot}$), these questions become even more fundamental.%: to build objects with masses below the hydrogen-burning mass limit ($\leq$ 0.072M$_{\odot}$), a high-density phase is necessary for the gravitational fragmentation to create very small Jeans-unstable cores. 

%One of the main open questions in the theory of star formation is: How do brown dwarfs and free-floating planets form. To build these objects with masses below the hydrogen-burning mass limit ($\leq$ 0.073M$_{\odot}$), a high-density phase is necessary for the gravitational fragmentation to create very small Jeans-unstable cores. 
Several scenarios have been proposed to prevent a substellar core in a dense environment from accreting to stellar mass (e.g., dynamical interactions, \citealt{Reipurth01,Umbreit05}, disk fragmentation, \citealt{Goodwin07, Stamatellos07}, or photoevaporation, \citealt{Whitworth04}). Alternatively, brown dwarfs and free-floating planets (with masses below the deuterium-burning mass limit, $\leq$ 13$M_{Jup}$) could form in an isolated mode by direct collapse. This could be possible either introducing turbulence so that the Jeans mass decreases in the first place \citep{Padoan02, Hennebelle08} or from a filament collapse (e.g., \citealt{Inutsuka92}) forming low-mass cores that experience high self-erosion in outflows \citep{Machida09}.
%{\bf need to add Stamatello's, Chabrier, etc work!!}

In order to test these models, observational constraints need to be placed on the main features of star formation (disk properties / morphology, accretion, outflows, etc.) toward the lowest possible masses, so that conclusions can be drawn from the behaviors of those properties with the mass of the central object.
%{\bf CHANGE!!! self-plag} A key to understanding star and brown dwarf formation is to observationally define the minimum mass that the canonical star formation process can produce by detecting and exploring the main features characteristic of star formation, such as disks, accretion, and outflows, for very low-mass objects.

Young brown dwarfs were shown to have substantial circumstellar material based on near-infrared (IR; e.g. \citealt{Oasa99,Muench00}), mid-IR (ISO \& $Spitzer$, e.g., \citealt{Natta01,Natta02,Apai05, Barrado07,Luhman08,Bayo12}), far-IR ($Herschel$, e.g., \citealt{Harvey12a,Harvey12b, Alves13, Joergens13,Liu15}) and single-dish millimeter continuum photometry (e.g., \citealt{Klein03, Scholz06, Mohanty13}). %, and more recently \citealt{Ricci14}). 
Only a few years ago, millimeter interferometers became more sensitive, and the disk around 2MJ0444 (M7.25, $\sim$0.05$M_{\odot}$) was the first of its kind to be spatially resolved at 1.3mm (CARMA observations by \citealt{Ricci13}) with an estimated disk radius of 15--30 au. 
%It was shown that disks of brown dwarfs exhibit the first steps of planet formation, such as grain growth, crystallization, and settling (e.g., Apai et al. 2005; Pascucci et al 2009). 
Even more recently, ALMA allowed several groups to perform small surveys in the substellar domain including intermediate-to-high-mass brown dwarfs \citep{Ricci14, Daemgen16, Pascucci16, Testi16, vanderPlas16}. Measurements of low spectral indices of a handful objects showed that dust grains have grown to millimeter sizes even in these very low-mass environments (e.g. \citealt{Ricci13}), challenging models of planetesimal formation, which predict that dust growth is more limited by radial drift than in disks around stellar objects \citep{Pinilla13}. Hints of disk truncation in the substellar domain were presented \citep{Testi16}, as well as different estimates of the temperature of the dust in substellar disks and an evolution \citep{Andrews13,vanderPlas16}, within the first 10 Myr, of the $M_{\rm dust}$--$M_{*}$ relationship (pointing again toward grain growth, drift, and fragmentation). 

Moving down in mass, the lowest-mass isolated objects found to harbor a disk are, to the best of our knowledge, Proplyd 133-353 ($\leq13\rm{M_{\rm Jup}}$, \citealt{Fang16}), Cha 1109-7734 ($\sim 8\rm{M_{\rm Jup}}$, \citealt{Luhman05}), J02265658-5327032 ($\sim13\rm{M_{\rm Jup}}$; \citealt{Boucher16}), and OTS44 ($\sim 12\rm{M_{\rm Jup}}$; \citealt{Joergens13}), with central object masses well below what has been studied with ALMA until now.

In this Letter we present the first millimeter detection of one of these four extremely low-mass objects, OTS44. In Section \ref{sec:OTS44intro}, we describe in more detail the target; in Section \ref{sec:ALMA}, we describe our observations; in Section \ref{sec:results}, we present our disk estimate and the comparison with the literature; and in Section~\ref{sec:conclusions}, we present our conclusions. 
%Figure 1 displays the spectral energy distribution (SED) from the optical to the far-IR of the free-floating planet OTS 44. ALMA early science observations (Cycle0) of the very low-mass star $\rho$ Oph 102 (M5.5, $\sim$0.13 M$_{\odot}$) allowed the first ALMA detection of the disk around an object close to the substellar border in 1-3mm continuum and CO (J=3-2) emission (Ricci et al. 2012). While $\rho$ Oph 102 is frequently claimed to be a brown dwarf with mass of $\sim$0.06 M$_{\odot}$, an optical spectrum shows that it is rather a very low-mass star (M5.5, $\sim$0.13 M$_{\odot}$, K. Luhman, pers. comm.). 

\section{OTS44}\label{sec:OTS44intro}
%OTS44 is one of only three free-floating planets known to have a disk. 
OTS44 is the object with the latest spectral type in the Chamaeleon I (Cha I) star forming region (M9.5) with a mass below or close to the planetary border (6-17$\rm{M_{\rm Jup}}$, \citealt{Luhman05b,Bonnefoy14}). First evidence for a disk around OTS44 came from mid- and far-IR excess emission detected with $Spitzer$ and $Herschel$ \citep{Luhman05b, Harvey12a, Harvey12b}. In addition, we observed OTS44 with VLT/SINFONI and detected strong, broad, and variable Pa $\beta$ emission, which is evidence for active disk accretion in the planetary regime with a relatively high mass-accretion rate (8$\times$10$^{-12}$ M$_{\odot}$ yr$^{-1}$; \citealt{Joergens13}).

We recently determined the properties of the disk of OTS44 \citep{Joergens13, Liu15} through radiative transfer modeling of its spectral energy distribution (SED) from the optical to the far-IR applying the radiative transfer code MC3D \citep{Wolf03} and a Bayesian analysis. The disk model that fitted the mid- and far-IR data best was that of a highly flared disk with a dust mass of 0.17$\rm{M_{\oplus}}$. However, our far-IR $Herschel$ measurements (a detection at 70$\mu$m and an upper limit at 160$\mu$m) are insensitive to millimeter-sized grains, which prevented us from concluding about the presence of large grains (a maximum grain size of 100$\mu$m was assumed in \citealt{Liu15}), with this potentially leading to an underestimation of the disk dust mass. In this Letter we report the first ALMA detection of the disk of a planetary-mass object, providing a more robust estimate of its mass and supporting the idea that the value obtained in \cite{Liu15} was indeed an underestimation.% the latter underestimation.
%{\bf comment on the discrepancies in the M24 data!}

%The exquisite sensitivity of ALMA at mm-wavelengths is ideally suited to accurately measure the disk dust mass of OTS 44 at those wavelength where the disk is optically thin and cold mm-sized grains can be detected. Millimeter surveys of young stars find a ratio of the disk-to-central-mass of about 10?2 (e.g., Williams \& Cieza 2011). This is displayed in Fig.2, where we also show our results for OTS44, which is the object with the lowest central mass for which the disk mass is estimated. It is of prime importance to refine the disk dust mass estimate of OTS 44 through mm-observations with ALMA. This will probe if this trend observed for stars extends to planetary mass objects.

\section{ALMA data}\label{sec:ALMA}

ALMA Cycle 3 Band 6 continuum data were obtained as part of the program {\it 2015.1.00243.S}. %The original goal was to also obtain Band 8 (412 GHz) data, but only one science goal was executed. 
Four spectral windows (centered at 224, 226, 240, and 242 GHz and each one with $\sim$1.9 GHz bandwidth) were defined to be collapsed in a single ``broadband" continuum image.  

The data processing was performed with CASA \citep{McMullin07}, following the standard steps starting from the measurement sets: visual inspection of the performance of antennas and scans, flagging corrupted or useless data (using solutions derived from the water vapor radiometer), correcting in bandpass, flux and phase, further flagging (shadowing, spectral window edges anomalies, etc.), deriving the bandpass solution per spectral window, and creating the cube from the calibrated data (the derived flux uncertainty is $\sim$8\%, but, based on the ALMA documentation, we assumed a more conservative 10\% value from now on). 

In addition to these general steps, we tried to apply self-calibration to improve the extended flux recovery but the signal-to-noise ratio (S/N) of the data was not good enough. Finally, we tried to bin our data in two spectral windows as separated as possible to obtain a spectral index, but once again, the S/N of the data was not good enough for this purpose.

\begin{figure}[ht!]
\gridline{\fig{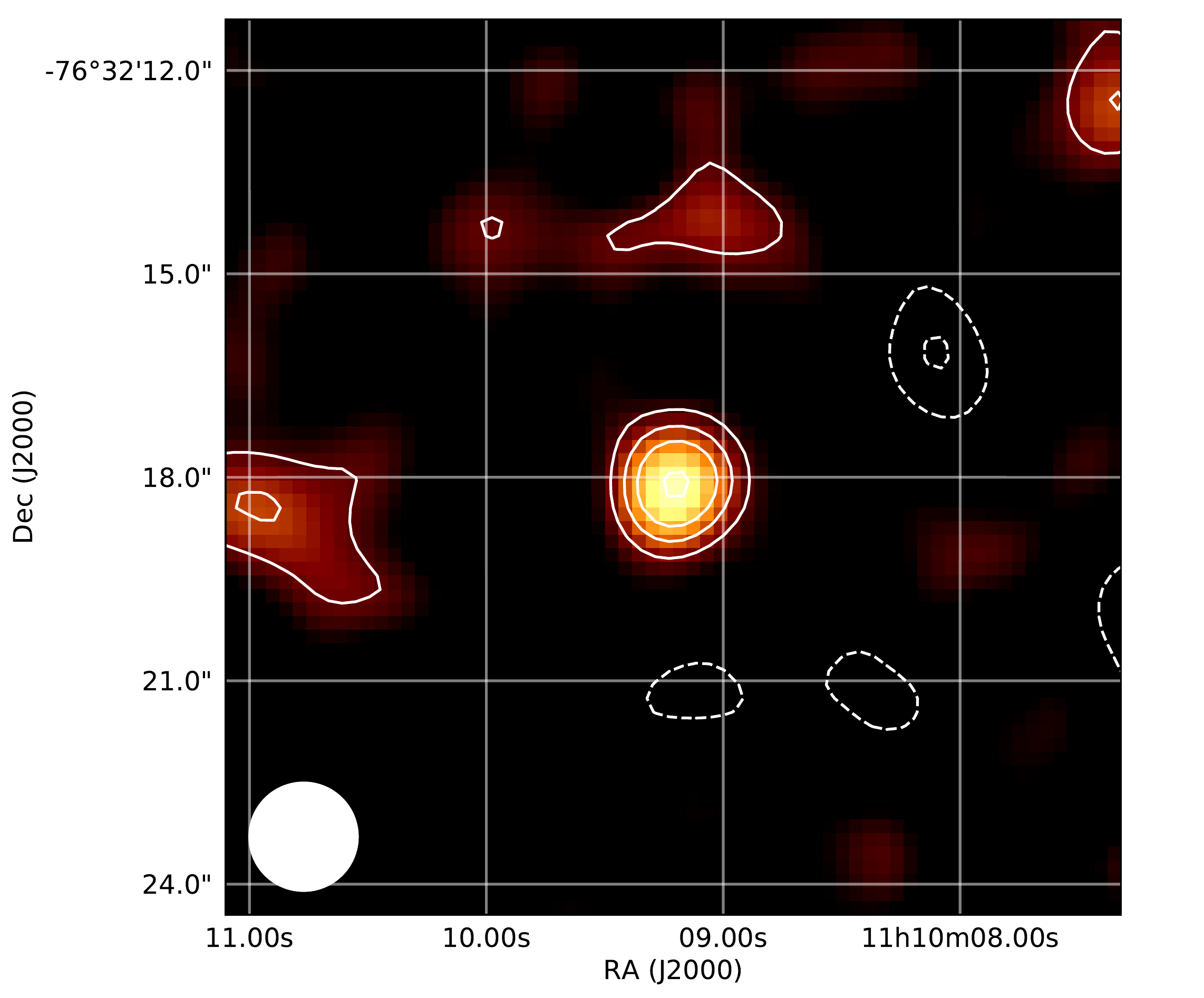}{0.5\textwidth}{\vspace{-0.1cm}}}
%\plotone{OTS44_ALMA_detection.pdf}
%\vspace{-0.1cm}
\caption{Frequency collapsed ALMA Band 6 data of OTS44. The 1.6" $\times$ 1.6" beam is displayed in the lower left corner. Solid-line white contours highlight regions with 3, 5, 7 and 10 times the RMS of the data (9.8$\mu$Jy/beam). Dashed-line white contours highlight $-3\times$RMS and $-5\times$RMS regions, and there are no data at the $-7$ and $-10$ RMS levels.\label{fig:detection}}
\end{figure}

The final CLEAN, primary beam corrected image (natural weighting) has a beam of 1.6" $\times$ 1.6", a RMS of 9.8$\mu$Jy/beam, a central reference frequency of 233 GHz, and a frequency range covered (not continuously) of 20 GHz. We detected OTS44 as a point-like source with a peak flux value 0.101$\pm$0.01mJy (see Fig.~\ref{fig:detection}).%at a $\sim$10$\sigma$ level. 
%OTS44: 0.101mJy
%Self-calibration not working, inspection of the channels no gas indeed.
%beam 1.5 $\times$ 1.4"
%Standard final reduction: RMS 9.8$\mu$Jy/beam
%LAS 30", so for sure no emission lost

\section{Results and discussion: dust disk mass}\label{sec:results}

The most direct quantity that we can derive from our data is the dust mass of the disk. In this section, we provide such estimates and compare them with the available literature.%For this, we follow two approaches and perform comparisons with the available literature data:

%From Testi: "Brown dwarf disks with ALMA: Evidence for truncated dust disks in Ophiuchus"
%From Gerit and Pascucci "just" that are suveys exploring the Mdust-Mstar relation showing hints of deviation from the stellar one and evolution in the first 10 Myrs (pointing again toward grain growth, drift, and fragmentation)
%Do not really know what to say here about daemgen, maybe comment for ll of them on Tdust, but probably that is better for the results section
%{\bf add a couple of lines on the Tdust dependence found in van der Plas and the regime change from Pascucci16}
%The observation of large grains in brown dwarf disks is a challenge for models of dust evolution in proto-planetary disks. Both radial drift and fragmentation of solids, which are known problems in the formation of planetesimals in T-Tauri disks, are amplified for the physical conditions in brown dwarf disks (Pinilla et al. 2013). Characterizing proto-planetary disks around the lowest mass objects provides key information for planet formation models.

\subsection{Disk mass via analytical prescription}\label{sec:analytical}

In order to estimate the disk dust mass from millimeter data, we assume that the emission is optically thin and isothermal at temperature T$_{\rm dust}$, and therefore

%$ \log M_{dust} = \log F_{\nu} + 2\log d - \log \kappa_{\nu} - \log B_{\nu}(T_{dust})$
$$ M_{dust} = \frac{F_{\nu} \times d^2}{\kappa_{\nu} \times B_{\nu}(T_{dust})}$$

where $F_{\nu}$ is the flux density, $d$ is the distance to Cha I (160 pc is assumed; \citealt{Whittet97}), $\kappa_{\nu}$ is the mass absorption coefficient, and $B_{\nu}$ is the Planck function of temperature $T_{\rm dust}$ at the observed frequency. 

%(2.3 opacity is for $\nu$ = 230GHz; assuming a dependence $\nu^{0.4}$ then for $\nu$ 338Ghz -> 2.68),
We have adopted a $\kappa_{\nu}$ value of 2.3 cm$^2$ g$^{-1}$ at 230 GHz with a frequency dependence of $\nu^{0.4}$, the same as in \cite{Andrews13} for Taurus, and more recently by \cite{Pascucci16} for Cha I. For consistency, we have adapted all M$_{\rm dust}$ values from the literature to be compatible with this assumption.

%{\bf MENTION HOW I HAVE PUT ALL EXPRESSIONS COMPATIBLE (CHANGING THE TESTI OPACITY, FOR EXAMPLE}

The remaining strong assumption rests on the choice of T$_{\rm dust}$. %, especially at the very low-mass end, where the luminosity of the central object is extremely low.
A typical value used for T$_{\rm dust}$ is 20 K, but \cite{Andrews13} showed that this temperature scales significantly with the luminosity of the central object,  proposing the relation: $T_{\rm dust} = 25 \times (L_*/L_{\odot})^{1/4}K$. However, this empirical relation yields a temperature of 5.5 K for OTS44 (assuming a luminosity of 0.0024 $L_{\odot}$; \citealt{Joergens13}), which is unrealistic given the higher temperatures reported merely by heating due to the interstellar radiation field (IRF; \citealt{Draine11}). Very recently,  \cite{vanderPlas16} revised this dependence as $T_{\rm dust} = 22 \times (L_*/L_{\odot})^{0.16}K$, which translates to a dust temperature of 8.4 K for OTS44 (with the same $L_{\odot}$ as before).

For completeness we have considered four different temperatures: the ``classical" 20 K, the 8.4 K obtained assuming the relationship from \cite{vanderPlas16}, that of IRF (7.5 K, \citealt{Draine11}), and the 5.5 K derived following the relationship from \cite{Andrews13}. The corresponding estimates for the dust mass of the disk of OTS44 are: 0.07, 0.27, 0.33, and 0.63 M$_{\oplus}$, respectively.

\begin{figure*}[ht!]
\gridline{\fig{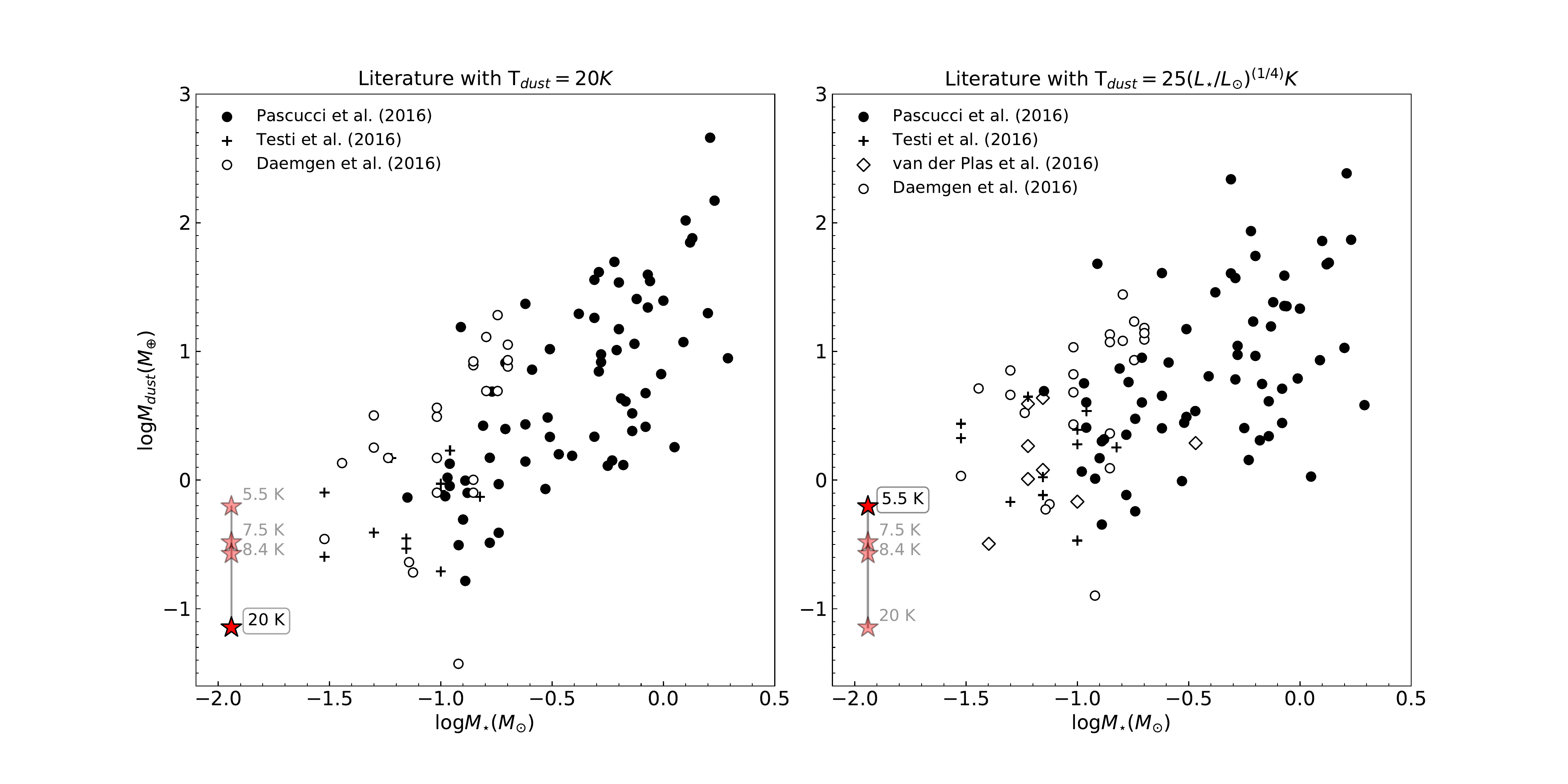}{1.\textwidth}{}}
%\plotone{bothfigs.pdf}
\vspace{-1cm}\caption{Illustration of the four (five-point red stars) possible determinations of the dust mass of the disk of OTS44 depending on the adopted temperature, in an $M_{*} - M_{disk}$ diagram, along with comparisons with other substellar objects reported in the literature. In each panel, we outline with a box the corresponding value of T$_{\rm dust}$ assumed for OTS44.}
\label{fig:comparisons}
\end{figure*}

In Fig.~\ref{fig:comparisons}, we show how our four estimates of the dust mass in the disk around OTS44 compare to literature values for stellar and substellar objects in the literature from \cite{Pascucci16, vanderPlas16, Daemgen16, Testi16}. We note that we are only comparing values derived from interferometric measurements since those derived from single-dish observations tend to be higher pointing toward a contamination of the measurement by the molecular cloud.

The estimates for the different samples shown in Fig.~\ref{fig:comparisons}, are consistent with each other since we recalculated the dust masses using the same absorption coefficient (to rescale the values in \citealt{Testi16}). 
%In the case of the constant T$_{\rm dust}$, we recalculated the values of \cite{Testi16} using consistent mass absorption coefficient with the remaining samples. 
In the case where T$_{\rm dust}$ scales with the bolometric luminosity of the central object, the same caution was taken in the use of consistent opacities, and we adopted the stellar parameters from \cite{Manara14,Manara16} and \cite{Manara17}, for the \cite{Pascucci16} sample (Cha I objects, as OTS44), where those values are generally in agreement with \cite{Luhman04} or more recently \cite{Bayo17}.

The conclusion from these comparisons is that a ``very high" T$_{\rm dust}$ value of 20 K translates in an extremely low-mass disk that would fall in the lower envelope of the $M_{*} - M_{disk}$ relationship drawn by the literature data. However, 20 K is probably an unrealistic value for the dust temperature unless disks around these very low-mass central objects are much smaller and flared than those around low-mass stars, a trend challenged, for example, by \cite{Liu15}, but worthy of further scrutiny. On the other hand, values between 5.5 K and 8.4 K, yield disk masses compatible with the dispersion observed in the literature data (although 5.5 K is probably also unrealistic due to the fact that including an IRF already brings this value to 7.5 K), pushing the $M_{*} - M_{disk}$ correlation into, or at the border of, the planetary-mass domain.

\section{Conclusions}\label{sec:conclusions}
We have presented the first millimeter detection of the dusty disk around an isolated planetary-mass object. Taking into account the strong assumptions to derive its dust mass, and following different approaches to do so, the values we obtain are consistent with the log-log linear relation between $M_{*}$ and M$_{\rm dust}$, holding even at the planetary-mass domain.

However, these mass estimates are severely limited by assumptions on poorly constrained parameters such as the dust properties in disks around these extremely low-mass objects. In addition, crucial aspects such as grain growth cannot be probed with one-band millimeter observations. To tackle these questions, the ideal complementary data would include $\sim$200$\mu$m (unfortunately, most likely beyond the limits of ALMA Band 10, $\sim$869 GHz, capabilities) observations and extremely sensitive ALMA Band 3 observations. 

\acknowledgments

This Letter makes use of the following ALMA data: ADS/JAO.ALMA\#2015.1.00243.S. ALMA is a partnership of ESO (representing its member states), NSF (USA) and NINS (Japan), together with NRC (Canada), NSC and ASIAA (Taiwan), and KASI (Republic of Korea), in cooperation with the Republic of Chile. The Joint ALMA Observatory is operated by ESO, AUI/NRAO and NAOJ. A. Bayo acknowledges financial support from the Proyecto Fondecyt Iniciaci\'on 11140572, and I. de Gregorio for very useful discussions. A.N. acknowledges support from Science Foundation Ireland (grant 13/ERC/12907). Y.L. acknowledges support by NSFC grant 11503087, by the Natural Science Foundation of Jiangsu Province of China (grant No. BK20141046), and by the German Academic Exchange Service and the China Scholarship Council.
J.O. acknowledges the UV support to the MPIA-UV Max Planck Tandem Group. H.B. acknowledges support from the European Science Council under the Horizon 2020 framework program via the ERC Consolidator grant CSF-648505. P.P. acknowledges support by NASA through Hubble Fellowship grant HST-HF2-51380.001-A awarded by the Space Telescope Science Institute, which is operated by the Association of Universities for Research in Astronomy, Inc., for NASA, under contract NAS 5- 26555. This research made use of: APLpy, an open-source plotting package for Python hosted at http://aplpy.github.com; Astropy, a community-developed core Python package for Astronomy \citep{astropy}; NASA's Astrophysics Data System; the SIMBAD database, operated at CDS, Strasbourg, France; and the VizieR catalog access tool, CDS, Strasbourg, France \citep{vizier}.

%We thank all the people that have made this AASTeX what it is today.  This
%includes but not limited to Bob Hanisch, Chris Biemesderfer, Lee Brotzman,
%Pierre Landau, Arthur Ogawa, Maxim Markevitch, Alexey Vikhlinin and Amy
%Hendrickson. Also special thanks to David Hogg and Daniel Foreman-Mackey
%for the new "modern" style design. It is super cool!


%\vspace{5mm}
\facility{ALMA}
%\facilities{ALMA}


\software{astropy \citep{astropy} 
%          Cloudy \citep{2013RMxAA..49..137F}, 
%          SExtractor \citep{1996A&AS..117..393B}
          }

\bibliographystyle{aasjournal.bst}
%\bibliography{biblio}



\end{document}